\documentclass[a4paper,twocolumn,english,english,aps]{revtex4}
\usepackage[T1]{fontenc}
\usepackage[latin1]{inputenc}
\usepackage{babel}
\usepackage{graphics}

\makeatletter

\providecommand{\LyX}{L\kern-.1667em\lower.25em\hbox{Y}\kern-.125emX\@}

\makeatother
\begin{document}

\title{Baryon bias and structure formation in an accelerating universe}

\author{Luca Amendola \& Domenico Tocchini-Valentini}

\address{Osservatorio Astronomico di Roma, \\
 Viale Frascati 33, \\
 00040 Monte Porzio Catone (Roma), Italy\\
 \textit{amendola@coma.mporzio.astro.it, tocchini@oarhp1.rm.astro.it}}

\date{\today {}}

\begin{abstract}
In most models of dark energy the structure formation stops when the
accelerated expansion begins. In contrast, we show that the coupling
of dark energy to dark matter may induce the growth of perturbations
even in the accelerated regime. In particular, we show that this occurs
in the models proposed to solve the cosmic coincidence problem, in
which the ratio of dark energy to dark matter is constant. Depending
on the parameters, the growth may be much faster than in a standard
matter-dominated era. Moreover, if the dark energy couples only to
dark matter and not to baryons, as requested by the constraints imposed
by local gravity measurements, the baryon fluctuations develop a constant,
scale-independent, large-scale bias which is in principle directly
observable. We find that a lower limit to the baryon bias \( b>0.5 \)
requires the total effective parameter of state \( w_{e}=1+p/\rho  \)
to be larger than \( 0.6 \) while a limit \( b>0.73 \) would rule
out the model.
\end{abstract}
\maketitle

\section{Introduction}

The epoch of acceleration which the universe seems to be experiencing
\cite{rie} is commonly regarded as a barren ground for what concerns
structure formation. In fact, during an accelerated expansion gravity
is unable to win over the global expansion and the perturbations stop
growing. Mathematically, this is seen immediately from the equation
governing the evolution of the perturbations in the Newtonian approximation
in a flat universe:\begin{equation}
\label{eqstandard}
\delta _{c}''+\left( 1+\frac{H'}{H}\right) \delta _{c}'-\frac{3}{2}\Omega _{c}\delta _{c}=0
\end{equation}
where \( H=d\log a/d\tau  \) is the Hubble constant in a conformally
flat FRW metric \( ds^{2}=a^{2}(-d\tau ^{2}+\delta _{ij}dx^{i}dx^{j}) \),
the subscript \( c \) stands for cold dark matter (here we neglect
the baryons) and the prime represents derivation with respect to \( \alpha =\log a \)
. When the dark energy field responsible of the acceleration becomes
dominant, \( \Omega _{c}\to 0 \) and the dominant solution of Eq.
(\ref{eqstandard}) becomes \( \delta _{c}\sim  \) const. Only if
gravity can overcome the expansion the fluctuations are able to grow.
It appears then that to escape the sterility of the accelerated regime
is necessary to prevent the vanishing of \( \Omega _{c} \). 

As it has been shown in ref. \cite{wet95}, an epoch of acceleration
with a non-vanishing \( \Omega _{c} \) can be realized by coupling
dark matter to dark energy. In fact, a dark energy scalar field \( \phi  \)
governed by an exponential potential linearly coupled to dark matter
yields, in a certain region of the parameter space, an accelerated
expansion with a constant ratio \( \Omega _{c}/\Omega _{\phi } \)
and a constant parameter of state \( w_{\phi } \), referred to as
a stationary accelerated era. Similar models have been discussed in
\cite{pav, dalal}. In \cite{amtoc2} we showed that in fact the conditions
of constant \( \Omega _{\phi } \) and \( w_{\phi } \) uniquely determine
the potential and the coupling of the dark energy field. In this sense,
the model we discuss below is the simplest stationary model: any other
one must include at least another parameter to modulate the parameter
of state. The main motivation to consider a stationary dynamics is
that it would solve the cosmic coicidence problem \cite{zla} of the
near equivalence at the present of the dark energy and dark matter
densities \cite{pav, bm, dalal}. The stationarity in fact ensures
that the two components have an identical scaling with time, at least
from some time onward, regardless of the initial conditions. In \cite{amtoc}
it was shown that, by a suitable modulation of the coupling, structure
forms before the accelerated era. Further theoretical motivations
for coupled dark energy have been put forward in ref. \cite{gasp}.

As it will be shown below, the coupling has three distinct, but correlated,
effects on Eq. (\ref{eqstandard}): first, as mentioned, it gives
a constant non-zero \( \Omega _{c} \) in the accelerated regime;
second, adds to the {}``friction{}'' \( (1+H'/H)\delta _{c}' \)
an extra term which, in general, may be either positive or negative;
third, adds to the dynamical term \( -\frac{3}{2}\Omega _{c}\delta _{c} \)
a negative contribution that enhances the gravity pull.

The dark energy coupling is a new interaction that always adds to
gravity (see e.g. \cite{wet95, dam96}). The coupling to the baryons
is strongly constrained by the local gravity measurements, so that
we assume for simplicity that the baryons are in fact not explicitely
coupled to the dark energy as suggested in \cite{dam} and, in the
context of dark energy, in \cite{ame3, bm}(of course there remains
the gravitational coupling). This \emph{species-dependent coupling}
breaks the equivalence principle, but in a way that is locally unobservable.
However, we show that there is an effect which is observable on astrophysical
scales and that may be employed to put a severe constraint on the
model. In fact, the baryon perturbations grow in the linear regime
with a constant, scale-independent, large-scale bias with respect
to the dark matter perturbations, that is in principle observable.
Although the bias can be either larger or smaller than unity, we find
that all the accelerated models require \( b<1 \) i.e. baryons less
clustered than dark matter (sometimes denotes anti-bias). Such a \emph{baryon
bias} would be a direct signature of an explicit dark matter-dark
energy interaction, well distinguishable from most other hydrodynamical
mechanisms of bias (see e.g. \cite{kly}).

\section{Coupled dark energy}

Consider three components, a scalar field \( \phi  \), baryons and
CDM described by the energy-momentum tensors \( T_{\mu \nu (\phi )}, \)
\( T_{\mu \nu (b)} \)and \( T_{\mu \nu (c)} \), respectively. General
covariance requires the conservation of their sum, so that it is possible
to consider a coupling such that, for instance,\begin{eqnarray*}
T_{\nu (\phi );\mu }^{\mu } & = & \sqrt{2/3}\kappa \beta T_{(c)}\phi _{;\nu },\\
T_{\nu (c);\mu }^{\mu } & = & -\sqrt{2/3}\kappa \beta T_{(c)}\phi _{;\nu },
\end{eqnarray*}
 where \( \kappa ^{2}=8\pi G \), while the baryons are assumed uncoupled,
\( T_{\nu (b);\mu }^{\mu }=0 \) because local gravity constraints
indicate a baryon coupling \( \beta _{b}<0.01 \) \cite{wet95, dam96}.
Let us derive the background equations in the flat conformal FRW metric.
The equations for this model have been already described in \cite{amtoc},
in which a similar model (but with a variable coupling) was studied.
Here we summarize their properties, restricting ourselves to the case
in which radiation has already redshifted away. The conservation equations
for the field \( \phi  \), cold dark matter, and baryons, plus the
Einstein equation, are\begin{eqnarray}
\phi ''+(2+\frac{H'}{H})\phi '+a^{2}U_{,\phi } & = & -\sqrt{2/3}\kappa \beta a^{2}\rho _{c},\nonumber \\
\rho '_{c}+3\rho _{c} & = & \sqrt{2/3}\kappa \beta \rho _{c}\phi ',\nonumber \\
\rho '_{b}+3\rho _{b} & = & 0\nonumber \\
H'+\frac{H}{2}\left[ 1+\kappa ^{2}(\frac{1}{2}\phi '^{2}-\frac{a^{2}}{H^{2}}U)\right]  & = & 0\label{sys} 
\end{eqnarray}
 where \( U(\phi )=U_{0}e^{-\sqrt{2/3}\mu \kappa \phi } \) . The
coupling \( \beta  \) can be seen as the relative strength of the
dark matter-dark energy interaction with respect to the gravitational
force. The only parameters of our model are \( \beta  \) and \( \mu  \)
(the constant \( U_{0} \) can always be rescaled away by a redefinition
of \( \phi  \)). For \( \beta =\mu =0 \) we reduce to the standard
cosmological constant case, while for \( \beta =0 \) we recover the
Ferreira \& Joyce model \cite{fer}. As shown in ref. \cite{wet95},
the coupling we assume here can be  derived by a conformal transformation
of a Brans-Dicke model, which automatically leaves the radiation uncoupled.
To decouple the baryons one needs to consider a two-metric Brans-Dicke
Lagrangian as proposed in \cite{dam}. 

The system (\ref{sys}) is best studied in the new variables \cite{cop, ame3}
\( x=\kappa \phi '/\sqrt{6},\quad y=\frac{\kappa a}{H}\sqrt{U/3}, \)
and \( u=\frac{\kappa a}{H}\sqrt{\rho _{b}/3} \) . Then we obtain
\begin{eqnarray}
x^{\prime } & =-\frac{1}{2} & \left( 3-3x^{2}+3y^{2}\right) x-\mu y^{2}+\beta (1-x^{2}-y^{2}-u^{2}),\nonumber \\
y^{\prime } & = & \mu xy+\frac{1}{2}y\left( 3+3x^{2}-3y^{2}\right) ,\nonumber \\
u^{\prime } & = & \frac{1}{2}u\left( 3x^{2}-3y^{2}\right) .\label{sys2} \\
 & \nonumber 
\end{eqnarray}
The CDM energy density parameter is obviously \( \Omega _{c}=1-x^{2}-y^{2}-u^{2} \)
while we also have \( \Omega _{\phi }=x^{2}+y^{2}, \) and \( \Omega _{b}=u^{2} \).
The system is subject to the condition \( x^{2}+y^{2}+u^{2}\leq 1 \). 

The critical points of system (\ref{sys2}) are listed in Tab. I.
We denoted with \( w_{e}=1+p_{tot}/\rho _{tot}=1+x^{2}-y^{2} \) the
total parameter of state. On all critical points the scale factor
expansion is given by \( a\sim \tau ^{p/1-p}=t^{p} \), where \( p=2/(3w_{e}) \),
while each component scales as \( a^{-3w_{e}} \). In the table we
also denoted \( g\equiv 4\beta ^{2}+4\beta \mu +18 \), and we used
the subscripts \( b,c \) to denote the existence of baryons or matter,
respectively, beside dark energy. In the same table we report the
conditions of stability and acceleration of the critical points, denoting
\( \mu _{+}=(-\beta +\sqrt{18+\beta ^{2}})/2 \).  

\begin{table*}

\caption{Critical points.}

\begin{tabular}{ccccccccc}
\hline 
Point &
 \( x \)&
 \( y \)&
 \( u \)&
 \( \Omega _{\phi } \)&
 \( p \)&
\( w_{e} \)&
stability&
 acceleration\\
\hline
\( a \)&
 \( -\frac{\mu }{3} \)&
 \( \sqrt{1-\frac{\mu ^{2}}{9}} \)&
 0 &
 1 &
 \( \frac{3}{\mu ^{2}} \)&
\( \frac{2\mu ^{2}}{9} \)&
\( \mu <\mu _{+},\mu <\frac{3}{\sqrt{2}} \)&
 \( \mu <\sqrt{3} \)\\
 \( b_{c} \)&
 \( -\frac{3}{2\left( \mu +\beta \right) } \)&
 \( \frac{\sqrt{g-9}}{2\left| \mu +\beta \right| } \)&
 \( 0 \)&
 \( \frac{g}{4\left( \beta +\mu \right) ^{2}} \)&
 \( \frac{2}{3}\left( 1+\frac{\beta }{\mu }\right)  \)&
\( \frac{\mu }{\mu +\beta } \)&
\( \beta >0,\mu >\mu _{+} \)&
 \( \mu <2\beta  \)\\
 \( b_{b} \)&
 \( -\frac{3}{2\mu } \)&
 \( \frac{3}{2\left| \mu \right| } \)&
 \( \sqrt{1-\frac{9}{2\mu ^{2}}} \)&
 \( \frac{9}{2\mu ^{2}} \)&
 \( \frac{2}{3} \)&
\( 1 \)&
\( \beta <0,\mu >\frac{3}{\sqrt{2}} \)&
 never\\
 \( c_{c} \)&
 \( \frac{2}{3}\beta  \)&
 0 &
 0 &
 \( \frac{4}{9}\beta ^{2} \)&
 \( \frac{6}{4\beta ^{2}+9} \)&
\( 1+\frac{4\beta ^{2}}{9} \)&
unstable \( \forall \mu \, \beta  \)&
 never\\
 \( d \)&
 \( -1 \)&
 0 &
 0 &
 1 &
 \( 1/3 \)&
2&
unstable \( \forall \mu \, \beta  \)&
 never \\
 \( e \)&
 \( +1 \)&
 0 &
 0 &
 1 &
 \( 1/3 \)&
2&
unstable \( \forall \mu \, \beta  \)&
 never \\
 \( f_{b} \)&
 0 &
 0 &
 1 &
 0 &
 \( 2/3 \)&
1&
unstable \( \forall \mu \, \beta  \)&
 never \\
\hline
\end{tabular}

\end{table*}

As it can be seen, the attractor \( a \) can be accelerated but \( \Omega _{c}\to 0 \),
so that structure cannot grow, as in almost all models studied so
far. Therefore, from now on we focus our attention on the global attractor
\( b_{c} \), the only critical point that may be stationary (i.e.
\( \Omega _{c} \) and \( \Omega _{\phi } \) finite and constant)
and accelerated. On this attractor the two parameters \( \beta  \)
and \( \mu  \) are uniquely fixed by the observed amount of \( \Omega _{c} \)
and by the present acceleration parameter (or equivalently by \( w_{e}=\mu /(\mu +\beta ) \)
). For instance, \( \Omega _{c}=0.20 \) and \( w_{e}=0.23 \) gives
\( \mu =3 \) and \( \beta =10 \).

\section{Differential growth rate }

Definining the perturbation variables \( \delta =\delta \rho /\rho ,\quad \frac{\sqrt{6}}{\kappa }\varphi =\delta \phi ,\quad ik^{i}\delta u_{i}/a=\theta H, \)
the following conservation equations for CDM, baryons and scalar field
in the synchronous gauge for the wavenumber \( k \) are derived:

\begin{eqnarray}
\delta '_{c} & = & -\theta _{c}-\frac{1}{2}h'-2\beta \varphi ',\label{cdm1} \\
\theta '_{c} & = & -\left( 1+\frac{H'}{H}\right) \theta _{c}+2\beta \left( -\frac{k^{2}}{H^{2}}\varphi +\theta _{c}x\right) ,\label{cdm} \\
\delta '_{b} & = & -\theta _{b}-\frac{1}{2}h',\\
\theta '_{b} & = & -\left( 1+\frac{H'}{H}\right) \theta _{b},\\
\varphi '' & + & (2+\frac{H'}{H})\varphi '+\frac{k^{2}}{H^{2}}\varphi +\frac{1}{2}h'x+2\mu ^{2}y^{2}\varphi =\beta \Omega _{c}\delta _{c}\label{kgpert} 
\end{eqnarray}
 Moreover we obtain the \textbf{}Einstein equation \textbf{}\begin{equation}
\label{einst}
h''=-(1+\frac{H'}{H})h'-2\left( 12\varphi 'x-6\mu y^{2}\varphi \right) +3(\delta _{c}\Omega _{c}+\delta _{b}\Omega _{b}).
\end{equation}

Now, deriving the \( \delta '_{c} \) equation we obtain \begin{eqnarray}
 & \delta ''_{c}+\left( 1+\frac{H'}{H}-2\beta x\right) \delta _{c}'+(\frac{4\beta ^{2}}{3}-1)\frac{3}{2}\delta _{c}\Omega _{c}-\frac{3}{2}\delta _{b}\Omega _{b} & =\nonumber \\
 & -6\varphi y^{2}\mu +\left( 12+4\beta ^{2}\right) \varphi 'x & \nonumber \label{deltacsec} \\
 & +4\beta (\frac{1}{2}\varphi '+\frac{k^{2}}{H^{2}}\varphi +\frac{1}{2}h'x+\mu ^{2}y^{2}\varphi ).\label{deltacsec} 
\end{eqnarray}
In the small-scale Newtonian approximation we can take the limit \( k\to \infty  \).
In Eq. (\ref{kgpert}) this amounts to neglecting the derivatives
of \( \varphi  \) and the potential term \( \mu ^{2}y^{2}\varphi  \),
which gives\textbf{\begin{equation}
\label{phiappr}
\frac{k^{2}}{H^{2}}\varphi +\frac{1}{2}h'x\approx \beta \Omega _{c}\delta _{c}.
\end{equation}
}Substituting in Eq. (\ref{deltacsec}) and neglecting again \( \varphi ',\varphi '' \)
and the potential term, we obtain\begin{equation}
\label{deltacsimp}
\delta ''_{c}+\left( 1+\frac{H'}{H}-2\beta x\right) \delta _{c}'-\frac{3}{2}\gamma \delta _{c}\Omega _{c}-\frac{3}{2}\delta _{b}\Omega _{b}=0,
\end{equation}
where \( \gamma \equiv 1+4\beta ^{2}/3, \) and similarly for \( \delta _{b} \)\begin{equation}
\label{deltabsimp}
\delta ''_{b}+\left( 1+\frac{H'}{H}\right) \delta _{b}'-\frac{3}{2}(\delta _{c}\Omega _{c}+\delta _{b}\Omega _{b})=0.
\end{equation}
Eq. (\ref{deltacsimp}) corrects the equation given in ref. \cite{ame3},
which had a wrong sign (the error gives only a minor effect for the
small \( \beta  \) considered in those papers). In Eq. (\ref{deltacsimp})
the differences with respect to Eq. (\ref{eqstandard}) that we mentioned
in the introduction appear clearly: the friction term and the dynamical
term are modified, and the value of \( \Omega _{c} \) is constant
along the stationary attractor. On the stationary attractor \( \Omega _{b}\to 0 \)
and Eqs. (\ref{deltacsimp}) and (\ref{deltabsimp}) can be written
as \begin{eqnarray*}
\delta ''_{c}+\frac{1}{2}\left( 4-3w_{e}-4\beta x\right) \delta _{c}'-\frac{3}{2}\gamma \Omega _{c}\delta _{c} & = & 0\\
\delta ''_{b}+\frac{1}{2}\left( 4-3w_{e}\right) \delta _{b}'-\frac{3}{2}\Omega _{c}\delta _{c} & = & 0
\end{eqnarray*}
where \( x,w_{e} \) and \( \Omega _{c} \) are given in Table I as
functions of the fundamental parameters \( \mu ,\beta  \) for any
critical point. The solutions are \( \delta _{c}=a^{m_{\pm }} \)and
\( \delta _{b}=ba^{m_{\pm }} \) where\begin{eqnarray}
m_{\pm } & = & \frac{1}{4}\left[ -4+3w_{e}+4\beta x\pm \Delta \right] \\
b_{\pm } & = & 3\Omega _{c}/(3\gamma \Omega _{c}+4\beta xm_{\pm })\label{sol} 
\end{eqnarray}
where \( \Delta ^{2}=\left( 24\gamma \Omega _{c}+(-4+3w_{e}+4\beta x)^{2}\right)  \).
The constant \( b\equiv \delta _{b}/\delta _{c}\equiv b_{+} \) is
the bias factor of the growing solution \( m\equiv m_{+} \). The
scalar field solution is \( \varphi \approx (H_{0}a^{(p-1)/p}/k)^{2}\delta _{c}(\beta \Omega _{c}+mbx) \).
For small wavelengths \( \varphi  \) (which is proportional to \( \delta \rho _{\phi }/\rho _{\phi } \)
) is always much smaller than \( \delta _{c},\delta _{b} \) at the
present time (although it could outgrow the matter perturbations in
the future if \( p>1 \)).

The solutions \( m_{\pm },b_{\pm } \) apply to all the critical solutions
of Table I. Let us now focus on the stationary attractor \( b_{c} \).
For \( \beta =0 \) we recover the law \( m_{\pm }=\frac{1}{4}\left[ -1\pm (24\Omega _{c}+1)^{1/2}\right]  \)
that holds in the uncoupled exponential case of Ferreira \& Joyce
\cite{fer}. Four crucial properties of the solutions will be relevant
for what follows: first, the perturbations grow (i.e. \( m>0 \))
for all the parameters that make the stationary attractor stable;
second, the baryons are anti-biased (i.e. \( b<1 \)) for the parameters
that give acceleration; third, in the \( k\gg H \) limit (and in
the linear regime), the bias factor is scale independent and constant
in time; and fourth, the bias is independent of the initial conditions.
Numerical integrations of the full set of equations (\ref{cdm1}-\ref{einst})
that confirm and illustrate the dynamics are shown in Fig. 1. Notice
that, in the future, the perturbations will reenter the horizon because
of the acceleration, so that the subhorizon regime in which our solutions
are valid will not hold indefinitely.

The species-dependent coupling generates a biasing between the baryon
and the dark matter distributions. In contrast, the bias often discussed
in literature concerns the distribution only of baryons clustered
in luminous bodies \cite{kly}. To observe the baryon bias one should
take into account also the baryons not collapsed in galaxies, e.g
in Lyman-\( \alpha  \) clouds or in intracluster gas. A measure of
the biasing of the total baryon distributions is possible in principle
but is still largely undetermined. Even taking the extremely simplified
approach that the baryon bias coincides with the galaxy bias we are
confronted with the problem that the galaxy biasing depends on luminosity
and type \cite{pea}. So we considered only very broad limits to \( b \):
since the acceleration requires anti-bias, we assume \( 0.5<b<1 \).
For a comparison, the likelihood analysis of ref. \cite{web} gives
\( b\in (0.8,1.9) \) (for IRAS galaxies, at 99\% c.l.), but is restricted
to \( \Lambda  \)CDM models with primordial slope \( n=1 \);  other
estimations do not exclude anti-bias, and might even require it \cite{rines}.

In Fig. 2 we show all the various constraints. To summarize, they
are: \emph{a}) the present dark energy density \( 0.6<\Omega _{\phi }<0.8 \);
\emph{b}) the present acceleration (\( w_{e}<2/3 \), implying \( \beta >\mu /2 \));
\emph{c}) the baryon bias \( 1>b>0.5 \). On the stationary attractor
there is a mapping between the fundamental parameters \( \mu ,\, \beta  \)
and the observables \( w_{e},\Omega _{\phi } \), so one can plot
the constraints on either pair of variables. It turns out that these
conditions confine the parameters in the small dark shaded area, corresponding
to \begin{eqnarray}
w_{e}\in (0.59-0.67), &  & \\
or\quad \beta \in (1.1-1.4) & , & \mu \in (2.0-2.6).\label{constr} 
\end{eqnarray}
Therefore, the parameters of the stationary attractor are determined
to within 20\% roughly. It is actually remarkable that an allowed
region exists at all. The growth rate \( m \) is approximately 0.5
in this region. For \( b>0.73 \) the possibility of a stationary
accelerated attractor able to solve the coincidence problem would
be ruled out. If one considers the tighter limit \( w_{\phi }<0.4 \)
for the supernovae Ia given at two sigma in ref. \cite{dalal} for
stationary attractors the allowed region would be further reduced,
possibly requiring a lower \( b \) to survive.
\begin{figure}
{\centering \resizebox*{9cm}{!}{\includegraphics{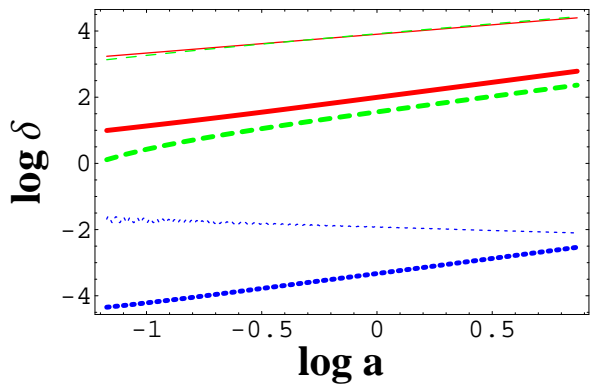}} \par}

\caption{Numerical evolution of the density contrast for a 100 Mpc/\protect\( h\protect \)
perturbation of dark matter (continuous lines), baryons (dashed lines)
and scalar field (dotted lines). Thick lines: \protect\( \beta ,\mu =1.5,3\protect \)
(or \protect\( \Omega _{\phi }=0.55,w_{e}=0.67\protect \)), resulting
in a bias \protect\( b\approx 0.3\protect \). Thin lines: \protect\( \beta ,\mu =0.25,3\protect \)
(or \protect\( \Omega _{\phi }=0.5,w_{e}=0.92\protect \)): here the
dark matter and baryon curves are almost indistinguishable since \protect\( b\approx 1\protect \).}
\end{figure}

\begin{figure}
{\centering \resizebox*{9cm}{!}{\includegraphics{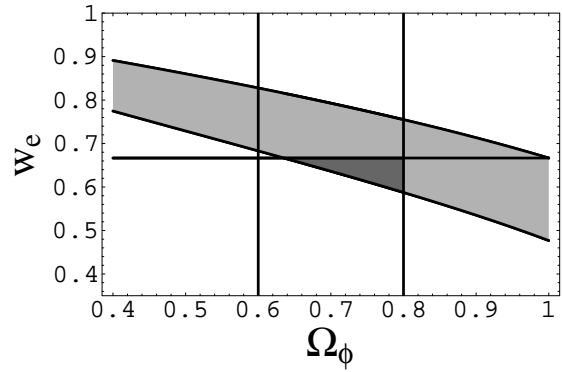}} \par}

\caption{Constraints on the stationary model: below the horizontal line the
expansion is accelerated; in the light grey region the bias is between
0.5 and 1; between the vertical lines \protect\( \Omega _{\phi }\protect \)
is within the observed range. The dark grey region is the surviving
parameter space.}
\end{figure}

\section{Conclusions}

We have shown that if the universe is experiencing a stationary epoch
capable of solving the cosmic coincidence problem then two novel features
arise in the standard picture of structure formation. First, a non-zero
\( \Omega _{c} \) during the accelerated regime allows structure
to grow; second, since the baryons have to be uncoupled (or very weakly
coupled), the growth is species-dependent, resulting in a constant
baryon bias independent of initial conditions. Although there are
no direct observations of the baryon bias, the trend is that more
massive objects are more biased with respect to the dark matter distribution,
so probably the galaxy bias is higher than the total baryon bias.
If this is correct, then \( b \) can be smaller than unity, as we
find to occur for all accelerated models. We find that the bias strongly
constrains the existence of a stationary epoch. Putting \( b>0.5 \),
and requiring \( 0.6<\Omega _{\phi }<0.8 \), we get that the two
free parameters \( \mu  \) and \( \beta  \) are fixed to a precision
of 20\% roughly, while the effective parameter of state \( w_{e} \)
is larger than 0.59. A higher bias or a lower \( w_{e} \) can easily
result in ruling out this class of stationary models. On the other
hand, the observation of a constant, scale-independent, large-scale
anti-bias would constitute a strong indication in favor of a dark
matter-dark energy coupling. 

The growth rate \( m \) is another observable quantity that can be
employed to test the stationarity, for instance estimating the evolution
of clustering with redshift. So far the uncertainties of this method
are overwhelming (see e.g. \cite{magli}) but future data should dramatically
improve its validity. The combined test of \( b \) and \( m \) will
be a very powerful test for the dark matter-dark energy interaction.
There is also the possibility to compare the resulting power spectrum
or cluster abundance with observations, although then we should know
exactly when the stationarity begun, and what dynamics preceded it
(see e.g. the variable coupling model of ref. \cite{amtoc}). 

Although we investigated only the simplest stationary model, in which
\( w_{e} \) is constant (a reasonable assumption over a small redshift
range), it is obvious to expect that a similar baryon bias develops
whenever there is a species-dependent coupling; this, in turn, is
requested to provide stationarity without conflicting with local gravity
experiments. Therefore, we conjecture that the baryon bias is a strong
test for all stationary dynamics.

\end{document}